\documentclass[aps,prd,twocolumn,nofootinbib,superscriptaddress]{revtex4-1}

\usepackage{amsfonts}
\usepackage{mathrsfs}
\usepackage{graphicx}
\usepackage{amsmath}
\usepackage{amssymb}
\usepackage{color}

\allowdisplaybreaks[4]

\usepackage[colorlinks=true,linkcolor=blue,citecolor=blue, urlcolor=blue]{hyperref}

\begin{document}

\title{Precise perturbative predictions from fixed-order calculations}

\author{Jiang Yan}
\email{yjiang@cqu.edu.cn}
\author{Zhi-Fei Wu}
\email{wuzf@cqu.edu.cn}
\address{Department of Physics, Chongqing Key Laboratory for Strongly Coupled Physics, Chongqing University, Chongqing 401331, P.R. China}

\author{Jian-Ming Shen}
\email{shenjm@hnu.edu.cn}
\address{School of Physics and Electronics, Hunan Provincial Key Laboratory of High-Energy Scale Physics and Applications, Hunan University, Changsha 410082, P.R. China}

\author{Xing-Gang Wu}
\email{wuxg@cqu.edu.cn}
\address{Department of Physics, Chongqing Key Laboratory for Strongly Coupled Physics, Chongqing University, Chongqing 401331, P.R. China}

\date{\today}

\begin{abstract}

The intrinsic conformality is a general property of the renormalizable gauge theory, which ensures the scale-invariance of a fixed-order series at each perturbative order. Following the idea of intrinsic conformality, we suggest a novel single-scale setting approach under the principle of maximum conformality (PMC) with the purpose of removing the conventional renormalization scheme-and-scale ambiguities. We call this newly suggested single-scale procedure as the PMC$_{\infty}$-s approach, in which an overall effective $\alpha_s$, and hence an overall effective scale is achieved by identifying the $\{\beta_0\}$-terms at each order. Its resultant conformal series is scale-invariant and satisfies all renormalization group requirements. The PMC$_{\infty}$-s approach is applicable to any perturbatively calculable observables, and its resultant perturbative series provides an accurate basis for estimating the contribution from the unknown higher-order (UHO) terms. Using the Higgs decays into two gluons up to five-loop QCD corrections as an example, we show how the PMC$_{\infty}$-s works, and we obtain $\Gamma_{\rm H}\big|_{\text{PMC}_{\infty}\text{-s}}^{\rm PAA} = 334.45^{+7.07}_{-7.03}~{\rm KeV}$ and $\Gamma_{\rm H}\big|_{\text{PMC}_{\infty}\text{-s}}^{\rm B.A.} = 334.45^{+6.34}_{-6.29}~{\rm KeV}$. Here the errors are squared averages of those mentioned in the body of the text. The Pad$\acute{e}$ approximation approach (PAA) and the Bayesian approach (B.A.) have been adopted to estimate the contributions from the UHO-terms. We also demonstrate that the PMC$_{\infty}$-s approach is equivalent to our previously suggested single-scale setting approach (PMCs), which also follows from the PMC but treats the $\{\beta_i\}$-terms from different point of view. Thus a proper using of the renormalization group equation can provide a solid way to solve the scale-setting problem.

\end{abstract}

\maketitle

\section{Introduction}

Due to its asymptotic freedom property~\cite{Gross:1973id, Politzer:1973fx}, the QCD running coupling ($\alpha_s$) becomes numerically small at short distances and allows perturbative calculations of cross sections or decay widths for high momentum transfer processes. To yield finite and reliable expression for perturbatively calculable observable, one needs to regulate and cancel the divergences that occur commonly in perturbative calculation. For the purpose, certain renormalization scheme and renormalization scale have to be introduced to finish the renormalization procedures. As the requirement of renormalization group invariance (RGI), a valid prediction for a physical observable must be independent to any choices of renormalization scheme and renormalization scale. However a truncated perturbation series does not automatically satisfy these requirements. The approach of using a guessed renormalization scale and choosing an arbitrary range for its uncertainty is conventional, which however leads to renormalization scheme-and-scale ambiguities and greatly depresses the predictive power of perturbative QCD (pQCD) theory.

One may hope to achieve a scheme-and-scale independent prediction by systematically computing higher-order enough QCD corrections. However, this hope is in conflict with the presence of the divergent $n! \alpha_s^n \beta^n_0$ renormalon series. And the complication of a higher-order loop calculation increases greatly with the number of loops as well as the number of external legs, even though great progresses on the loop calculation technology have been made in the literature in recent years. It is thus important to find a proper way to achieve precise information as much as possible from the known perturbative series. The renormalization scale-setting problem then becomes an important problem for achieving precise fixed-order pQCD predictions~\cite{Wu:2013ei}.

In the literature, the principle of maximum conformality (PMC)~\cite{Brodsky:2011ta, Brodsky:2011ig, Mojaza:2012mf, Brodsky:2012rj, Brodsky:2013vpa} has been suggested to eliminate conventional renormalization scale-and-scheme ambiguities. The $\alpha_s$-running behavior is governed by the renormalization group equation (RGE). Then the $\{\beta_i\}$-terms emerged in perturbative series can be inversely adopted for fixing the correct value of $\alpha_s$ for a physical observable. The key point of PMC is to fix the correct magnitude of $\alpha_s$ by using the RGE recursively. The PMC also provides a solid way to extend the well-known Brodsky-Lepage-Mackenzie (BLM) approach~\cite{Brodsky:1982gc} to all orders. It has been demonstrated that the PMC prediction is independent to any choice of renormalization scale and scheme~\cite{Wu:2019mky}, which is consistent with the fundamental renormalization group approach~\cite{Stueckelberg:1953dz, Peterman:1978tb, GellMann:1954fq, rge4, Wu:2014iba} and the self-consistency requirements of the renormalization group~\cite{Brodsky:2012ms}.

The PMC is initially introduced as a multi-scale-setting approach~\cite{Brodsky:2011ta, Brodsky:2011ig, Mojaza:2012mf, Brodsky:2012rj, Brodsky:2013vpa}, in which distinct PMC scales at each order are determined by using different categories of the $\{\beta_i\}$-terms emerged at the corresponding orders. Because of it's perturbative nature, the resultant PMC scale-invariant series still has two kinds of residual scale dependence due to unknown higher-order (UHO) terms~\cite{Zheng:2013uja}; i.e., the last perturbative terms of the PMC scales are unknown (\textit{first kind of residual scale dependence}), and the last perturbative terms of the pQCD approximant are also not fixed since there is no information to fix its PMC scale (\textit{second kind of residual scale dependence}). These two residual scale dependence shall be suppressed at high orders in $\alpha_s$ and/or from exponential suppression. However, if the convergence of the perturbative series of either the PMC scale or the pQCD approximant is weak, such residual scale dependence could be significant. Then as a step forward, it is helpful to find a proper way to further suppress the residual scale dependence of the PMC predictions.

It has been found that the intrinsic conformality (iCF), which ensures the scale-invariance at each order by only identifying the $\{\beta_0\}$-terms at each order, could be helpful to suppress the residual scale dependence. By taking the iCF into account, an infinity-order scale-setting approach that is called as the PMC$_{\infty}$ approach~\cite{DiGiustino:2020fbk} has been proposed. Ref.\cite{Chawdhry:2019uuv} suggests another alternative PMC multi-scale-setting procedure (PMCa), whose effective scales are fixed by requiring all the scale-dependent terms at each order vanish. It has been shown that the PMCa and the PMC$_\infty$ approaches are equivalent to each other~\cite{Huang:2021hzr}. This equivalence reflects the fact that the iCF requires the scale invariance of the pQCD series at each order, and vice versa. Some applications of the PMC$_{\infty}$ approach can be found in Refs.\cite{DiGiustino:2021nep, Gao:2021wjn}. It is however still a multi-scale-setting approach, even though the newly determined PMC scales at each orders are free of \textit{first kind of residual scale dependence}. It still has the \textit{second kind of residual scale dependence}, which may be sizable for lower-order predictions. Furthermore, it may meet very small scale problem at some specific orders~\cite{Huang:2021hzr}. Towards the goal of achieving minimum scale-dependent prediction, we shall adopt the idea of iCF and introduce a novel single-scale approach (named PMC$_\infty$-s) to set an overall effective scale of the process. We shall show that it can also fix the conventional renormalization scale ambiguity and provide further suppression to the residual scale dependence. The PMC$_\infty$-s series shall also be helpful to estimate the contribution from the UHO-terms.

The idea of single-scale setting approach has been initially suggested in Refs.\cite{Grunberg:1991ac, Brodsky:1995tb} by using the BLM approach, which extends the BLM to two-loop QCD corrections. Lately, it has been found that by transforming the $n_f$-series into $\{\beta_i\}$-series correctly with the help of RGE, an all-orders single-scale approach by using the PMC (named PMCs) can be achieved~\cite{Shen:2017pdu}. It has been shown that the PMCs approach can be served as a reliable substitute for the multi-scale PMC approach. And it does lead to more precise pQCD predictions with less residual scale dependence, cf. Refs.\cite{Wu:2019mky, Huang:2021hzr} and references therein. The PMCs also provides a self-consistent way to achieve precise $\alpha_s$ running behavior in both the perturbative and nonperturbative domains~\cite{Deur:2017cvd, Yu:2021yvw}. It is thus interesting to show what's the relation between the PMC$_\infty$-s and the PMCs approaches. Moreover, by using the scale-invariant PMC$_\infty$-s series, we shall adopt two approaches to estimate the contributions from the UHO-terms, which are important to extend the predictive power of the pQCD theory.

The remaining parts of the paper are organized as follows. We give the PMC$_\infty$-s approach in Sec.~\ref{CT}, a demonstration of its equivalence to the PMCs approach is also given. As an application, we apply the PMC$_\infty$-s approach to deal with the important Higgs decay channel, $H\to gg$, up to five-loop QCD corrections. Numerical results and discussions are given in Sec.~\ref{Example}. Sec.~\ref{Summary} is reserved for a summary.

\section{Calculation Technology}\label{CT}

The $\alpha_s$-running behavior is governed by the following standard RGE,
\begin{equation}
\beta(a_{s}(\mu_{r}))=\mu_{r}^2\frac{{\rm d} a_{s}(\mu_{r})} {{\rm d}\mu_{r}^2} = -a_{s}^2(\mu_{r}) \sum_{i=0}^\infty \beta_i a^{i}_{s}(\mu_{r}), \label{rge}
\end{equation}
where the right-hand-side is a perturbative expansion in terms of $a_s=\alpha_s/4\pi$. By using the RGE, one can fix the $\alpha_s$ value at any perturbative scale by using the measurements of high-energy observables that can fix the coupling at a given scale such as $M_Z$. The \{$\beta_{i}$\}-functions have been known up to five-loop level in the modified minimal-subtraction scheme ($\overline{\rm MS}$-scheme)~\cite{Gross:1973ju, Politzer:1974fr, Caswell:1974gg, Tarasov:1980au, Larin:1993tp, vanRitbergen:1997va, Chetyrkin:2004mf, Czakon:2004bu, Baikov:2016tgj}.

By using the Taylor expansion, one can derive a scale-displacement relation of the couplings at two different scales $\mu_1$ and $\mu_2$,
\begin{equation}
	a_s^{k}(\mu_2) = a_s^{k}(\mu_1) + \sum_{n=1}^\infty (-1)^{n}\Theta_{k}^{(n)}\left(a_s(\mu_1)\right) \ln^{n}\frac{\mu_1^2}{\mu_2^2} , \label{running1}
\end{equation}
where we have introduced the notation
\begin{equation}
	\Theta_{k}^{(n)}\left(a_s(\mu_1)\right)=\frac{1}{n!}\left. \frac{{\rm d}^n a_s^{k}(\mu_{r})}{({\rm d} \ln\mu_{r}^2)^n}\right|_{\mu_{r}=\mu_1}.
\end{equation}
We thus have
\begin{align*}
	\Theta_{k}^{(1)}(a_{s}(\mu_{r}))=&\,ka_{s}^{k-1}\beta,\\	\Theta_{k}^{(2)}(a_{s}(\mu_{r}))=&\,\frac{1}{2!}ka_{s}^{k-1}\beta\left[(k-1)\frac{\beta}{a_{s}}+\frac{{\rm d}\beta}{{\rm d} a_{s}}\right],\\	\Theta_{k}^{(3)}(a_{s}(\mu_{r}))=&\,\frac{1}{3!}ka_{s}^{k-1}\beta\left[(k-1)(k-2)\frac{\beta^{2}}{a_{s}^{2}}\right.\\
	&\left.+3(k-1)\frac{\beta}{a_{s}}\frac{{\rm d}\beta}{{\rm d} a_{s}}+\left(\frac{{\rm d}\beta}{{\rm d} a_{s}}\right)^{2}+\beta\frac{{\rm d}^{2}\beta}{{\rm d} a_{s}^{2}}\right],\\
	\cdots . &
\end{align*}

\subsection{The newly suggested PMC$_\infty$-s approach}

Generally, a pQCD calculable physical observable $\rho$ can be written as
\begin{equation}\label{eqstart}
\rho = \sum_{i=1}^{p}\mathcal{L}_{i}(\mu_r, Q)a_{s}^{n+i-1}(\mu_{r}),
\end{equation}
where $n$ is the $\alpha_s$-power of the leading-order terms, $\mu_{r}$ is the renormalization scale, $Q$ represents the kinematic scale at which the observable is measured or the typical momentum flow of the process, and
\begin{displaymath}
\mathcal{L}_{i}(\mu_r, Q) = \sum_{j=0}^{i-1} c_{i,j}(\mu_r, Q)n_{f}^{j}
\end{displaymath}
is the perturbative coefficients with $n_f$ being the number of active light-quark flavors. When the sum is over all perturbative terms, corresponding to $p\to\infty$, the physical observable shall be independent to any choice of $\mu_r$ due to the cancellation of $\mu_r$-dependence among all the perturbative terms. While for a fixed-order prediction, one needs to be careful about the scale-setting problem. One can further separate the perturbative coefficients $\mathcal{L}_{i}$ into the scale-invariant iCF ones $\mathcal{L}_{i,\rm IC}(Q)$ and the scale-dependent ones $\tilde{\mathcal{L}}_{i}(\mu_{r}, Q)$, i.e.
\begin{equation}\label{decoupling}
\mathcal{L}_{i}(\mu_{r}, Q)=\mathcal{L}_{i,\rm IC}(Q) +\tilde{\mathcal{L}}_{i}(\mu_{r}, Q).
\end{equation}
The first IC coefficient $\mathcal{L}_{1,\rm{IC}}$ equals to $\mathcal{L}_{1}$, the second IC coefficient $\mathcal{L}_{2,\rm{IC}}$ can be derived by setting $n_f\equiv {33}/{2}$ to the RGE-involved $n_f$-terms in $\mathcal{L}_{2}$ such that to remove the non-conformal contribution from the NLO-terms, $\mathcal{L}_{2,\rm{IC}}= \mathcal{L}_{2}|_{n_f={33}/{2}}$, and etc. The perturbative series of $\rho$ under different choices of scale can be related by using the RGE, e.g. its series at any other scale can be derived from its perturbative series at $\mu_r=Q$. Then Eq.\eqref{eqstart} can be reorganized as the following form,
\begin{widetext}
\begin{align}\label{pQCD expansion}
\rho =&\,\sum_{i=1}^{p}\mathcal{L}_{i}(Q, Q)a_{s}^{n+i-1}(\mu_{r}) +\sum_{i=1}^{p}\sum_{j=1}^{p}(-1)^{j}\mathcal{L}_{i}(Q, Q)\Theta_{n+i-1}^{(j)} \left(a_{s}(\mu_{r})\right)\ln^{j}\frac{\mu_{r}^{2}}{Q^{2}},
\end{align}
\end{widetext}
where the scale-dependent log-terms have been explicitly written. Applying the scale-displacement relation \eqref{running1} to Eq.\eqref{pQCD expansion}, we then obtain
\begin{widetext}
\begin{align}\label{rewritting expansion at Q*}
	\rho=&\,\sum_{i=1}^{p}\mathcal{L}_{i,\rm IC}(Q) a_{s}^{n+i-1}(Q_{*}) +\sum_{i=1}^{p}\tilde{\mathcal{L}}_{i}(Q, Q)a_{s}^{n+i-1}(Q_{*}) +\sum_{i=1}^{p}\sum_{j=1}^{p}(-1)^{j} \mathcal{L}_{i}(Q, Q) \Theta_{n+i-1}^{(j)}\left(a_{s}(Q_{*})\right)\ln^{j}\frac{Q_{*}^{2}}{Q^{2}},
\end{align}
\end{widetext}
which is free of $\mu_r$-dependence and is equivalent to set $\mu_r=Q_{*}$ in Eq.\eqref{pQCD expansion} directly. This shows that if the value of scale $Q_{*}$ is fixed by requiring all the non-conformal terms or explicitly the last two summing terms in the right-hand-side of Eq.\eqref{rewritting expansion at Q*} to vanish, we shall get the required scale-invariant pQCD series. This scale-setting procedure well satisfies the property of iCF. We call it as the PMC$_\infty$-s approach, which only needs to fix one overall effective scale $Q_{*}$ and results in the following scale-invariant series,
\begin{equation}
	\rho=\sum_{i=1}^{p}\mathcal{L}_{i,\rm IC}(Q)\ a_{s}^{n+i-1}(Q_{*}).
\end{equation}
It shows that the PMC$_\infty$-s approach solves the conventional $\mu_r$-dependence, and it also removes the \textit{second kind of residual scale dependence}. To determine the magnitude of $Q_*$, it is convenient to fix the value of $\ln Q_{*}^{2}/Q^{2}$, which can be expanded as a perturbative series,
\begin{align}\label{Expression of Qstar} \ln\frac{Q_{*}^{2}}{Q^{2}}=\sum_{k=0}^{p-2}S_{k}a_{s}^{k}(Q_{*}) =\sum_{k=0}^{n}F_{k}a_{s}^{k}(Q_{0}).
\end{align}
Due to the perturbative nature of $\ln Q_{*}^{2}/Q^{2}$, the PMC$_\infty$-s approach recovers the \textit{first residual scale dependence} occurred in PMC original multi-scale approach~\cite{Brodsky:2011ta, Brodsky:2011ig, Mojaza:2012mf, Brodsky:2012rj, Brodsky:2013vpa}, which however shall be highly suppressed, e.g. it is at the ${\cal O}(a_s^{n+p+1})$-order level for a given $(n+p)_{\rm th}$-order fixed-order series. More specifically, the \textit{first kind of residual scale dependence} is at the order of $\left(\sum\limits_{i=1}^{p} \mathcal{L}_{i,\rm IC}(Q) a_s^{i-1}\right){\cal O}(a_s^{n+p+1})$. The first equality of Eq.(\ref{Expression of Qstar}) stands for the exact solution which can be solved numerically. Practically, one can expand the series over a critical coupling $a_s(Q_0\sim Q)$ up to the required order such that the difference between the expansion over $a_s(Q_0)$ and the exact one is less than $1\%$ or $0.1\%$. Then, if the right-hand-side series is convergent or the scale $Q_0$ is large enough such that the strong coupling $\alpha_s(Q_0)\ll 1$, we can adopt the second equality as the basis to do the analysis~\cite{Shen:2017pdu}. The coefficients $S_{i}$ or $F_{i}$ up to next-to-$\cdots$-to-leading-log (N$^{n}$LL) accuracy can be determined by a N$^{n+1}$LO pQCD calculation. The expressions of the iCF coefficients $\mathcal{L}_{i,\rm IC}$, $S_{i}$ and $F_{i}$ up to five-loop QCD corrections, and the relations between $S_{i}$ and $F_{i}$ are given in the Appendix for convenience. In addition to the standard fixing of iCF coefficients $\mathcal{L}_{i,\rm IC}(Q)$, the above solution also implies another way of setting the iCF coefficients, e.g.
\begin{widetext}
\begin{align}\label{relation}
 &\sum_{i=1}^{p}\left(\mathcal{L}_{i,\rm IC}(Q) -\mathcal{L}_{i}(Q, Q)\right) a_{s}^{n+i-1}(Q_{*}) = \sum_{i=1}^{p}\sum_{j=1}^{p}(-1)^{j} \mathcal{L}_{i}(Q, Q) \Theta_{n+i-1}^{(j)}\left(a_{s}(Q_{*})\right) \left(\sum_{k=0}^{p-2}S_{k}a_{s}^{k}(Q_{*})\right)^{j}.
\end{align}
\end{widetext}

\subsection{Equivalence of the PMC$_\infty$-s and the PMC-s approaches}

In the literature, the PMCs approach also provides an all-orders single-scale approach to fix the conventional renormalization scale ambiguity~\cite{Shen:2017pdu}. Though the starting point is quite different, in the following, we shall show that the PMCs and the PMC$_\infty$-s approaches are equivalent to each order. For the purpose, we first give a brief introduction of the PMCs approach.

The PMCs approach adopts the standard RGE and the general QCD degeneracy relations among different orders~\cite{Bi:2015wea} to transform the RG-involved $n_f$-series into the $\{\beta_i\}$-series. Then Eq.(\ref{eqstart}) can be rewritten as the following form,
\begin{widetext}
\begin{align}\label{pQCD approximant}		
	\rho=&\,r_{1,0}a_{s}^{n}(\mu_{r})+\left[r_{2,0}+n\beta_{0}r_{2,1}\right]a_{s}^{n+1}(\mu_{r}) +\bigg[r_{3,0}+n\beta_{1}r_{2,1}+(n+1)\beta_{0}r_{3,1} +\frac{n(n+1)}{2}\beta_{0}^{2}r_{3,2}\bigg]a_{s}^{n+2}(\mu_{r})\notag\\		&+\bigg[r_{4,0}+n\beta_{2}r_{2,1}+(n+1)\beta_{1}r_{3,1}+\frac{n(3+2n)}{2}\beta_{0}\beta_{1}r_{3,2} +(n+2)\beta_{0}r_{4,1}+\frac{(n+1)(n+2)}{2}\beta_{0}^{2}r_{4,2}\notag\\
	&+\frac{n(n+1)(n+2)}{3!}\beta_{0}^{3}r_{4,3}\bigg]a_{s}^{n+3}(\mu_{r}) +\bigg[r_{5,0}+n\beta_{3}r_{2,1}+(n+1)\beta_{2}r_{3,1} +\frac{n(n+2)}{2}\left(\beta_{1}^{2}+2\beta_{0}\beta_{2}\right)r_{3,2}\notag\\
	&+(n+2)\beta_{1}r_{4,1}+\frac{(n+1)(2n+5)}{2}\beta_{0}\beta_{1}r_{4,2} +\frac{n(3n^{2}+12n+11)}{6}\beta_{0}^{2}\beta_{1}r_{4,3}+(n+3)\beta_{0}r_{5,1}\notag\\
	&+\frac{(n+2)(n+3)}{2}\beta_{0}^{2}r_{5,2}+\frac{(n+1)(n+2)(n+3)}{6}\beta_{0}^{3}r_{5,3} +\frac{n(n+1)(n+2)(n+3)}{24}\beta_{0}^{4}r_{5,4}\bigg]a_{s}^{n+4}(\mu_{r})+\cdots.
\end{align}
\end{widetext}
The  renormalization scale dependent perturbative coefficients $r_{i,j}$ can be redefined as
\begin{equation}\label{rij-hatrij}
	r_{i,j}=\sum_{k=0}^{j}C_{j}^{k}\hat{r}_{i-k,j-k}\ln^{k}\frac{\mu_{r}^{2}}{Q^{2}},
\end{equation}
where $C_{j}^{k}=j!/\left(k!(j-k)!\right)$ are combination coefficients, and $\hat{r}_{i,j}=r_{i,j}|_{\mu_{r}=Q}$. Specially, we have $\hat{r}_{i,0}=r_{i,0}$. Then following the standard procedures of the PMCs, all the non-conformal $\{\beta_i\}$-terms are resummed to fix an overall effective coupling $a_s(Q_*)$, and the N$^{p-1}$LO-order pQCD prediction with $\alpha^{n}_{s}$ for leading-order terms \eqref{pQCD approximant} changes to the following conformal series,
\begin{equation}
\rho=\sum_{i=1}^{p} \hat{r}_{i,0} a_{s}^{n+i-1}(Q_{*}). \label{pmcs-series}
\end{equation}
Similarly, to determine the magnitude of $Q_*$, we can expand $\ln Q_{*}^{2}/Q^{2}$ as a power series over $a_{s}(Q_{*})$~\cite{Wu:2019mky},
\begin{equation}
	\ln\frac{Q_{*}^{2}}{Q^{2}}=\sum_{k=0}^{p-2} T_{k} a^{k}_{s}(Q_*),
\end{equation}
which can be fixed up to N$^{p-2}$LL-accuracy for a given N$^{p-1}$LO-order pQCD series. And after a careful calculation, we observe that
\begin{equation}
 \hat{r}_{i,0}=\mathcal{L}_{i,\rm IC},\;\;(i=1,2,\cdots,p)
\end{equation}
and
\begin{equation}
	T_{k}=S_{k},\;\;(k=0,1,\cdots,p-2).
\end{equation}
Thus the predictions using the PMC$_{\infty}$-s approach and the PMCs approach are overlap with each other. The explicit equivalence of the PMC$_{\infty}$-s and the PMCs formulas up to the case of $p=5$ can be achieved by comparing Eqs.(\ref{L1}-\ref{T2}) with the PMCs formulas given in Refs.\cite{Wu:2019mky, Shen:2017pdu}. The PMCs conformal series (\ref{pmcs-series}) satisfies the renormalization group invariance~\cite{Wu:2018cmb} and its prediction is scheme-and-scale independent.

\section{A new analysis of the Higgs boson decays into two gluons using the PMC single-scale approach}\label{Example}

The $H\to gg$ decay plays an important role in Higgs phenomenology. In this section, we take it as an explicit example to show that by using the PMC single-scale approach (PMC$_\infty$-s), a more accurate pQCD prediction can be achieved. Its total decay width $\Gamma_{\rm H}$ up to N$^4$LO-level takes the form
\begin{align}\label{Convseries}
	\Gamma_{\rm H}=\sum_{i=1}^{5}\Gamma_{i}=\frac{M_{H}^{3}G_{F}}{36\sqrt{2}\pi} \sum_{i=1}^{5}C_{i}(\mu_{r}) a_{s}^{i+1}(\mu_{r}),
\end{align}
where $i=(1,\cdots,5)$, which represents the LO-terms, the NLO-terms, ..., and the N$^4$LO-terms, respectively. The Fermi constant $G_{F}=1.16638\times10^{-5}$ GeV$^{-2}$. The perturbative coefficients $C_{i\in[1,5]}(M_{H})$ under the $\overline{\rm MS}$-scheme have been given in Refs.\cite{Inami:1982xt, Djouadi:1991tka, Graudenz:1992pv, Dawson:1993qf, Spira:1995rr, Dawson:1991au, Chetyrkin:1997iv, Chetyrkin:1997un, Baikov:2006ch, Herzog:2017ohr}, which can be conveniently transformed to mMOM-scheme ones under the Landau gauge~\cite{Zeng:2015gha, Zeng:2018jzf}~\footnote{To apply the PMCs, it is helpful to transform the $\overline{\rm MS}$-series into MOM-one with the help of the relation~\cite{Celmaster:1979km, Celmaster:1979dm, Celmaster:1979xr, Celmaster:1980ji}, since the MOM scheme is physical scheme and avoids the ambiguities of transforming the $n_{f}$-terms into the required $\{\beta_i\}$-terms. The mMOM-scheme is gauge dependent~\cite{Zeng:2020lwi}, and for definiteness, we adopt the Landau gauge to do our analysis.}. To do the numerical calculation, we take~\cite{ParticleDataGroup:2020ssz}, $M_{H}=125.25\pm0.17~{\rm GeV}$, $m_{t}=172.76\pm0.30~{\rm GeV}$ and $\alpha_{s}(M_{Z})=0.1179\pm0.0009$.

\subsection{Basic properties of the total decay width $\Gamma_{\rm H}$ up to N$^4$LO QCD corrections}
\label{Decay width}

\begin{figure}[htb]
	\centering
	\includegraphics[width=0.45\textwidth]{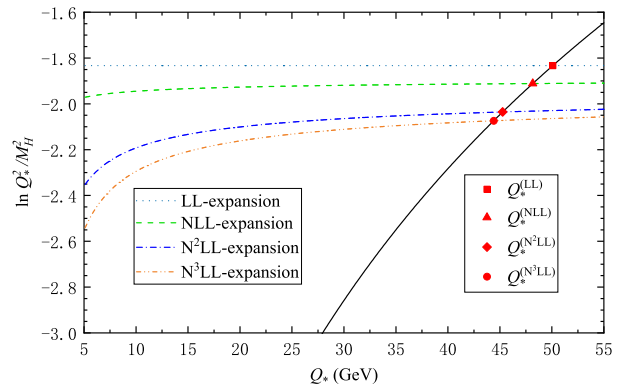}
	\caption{The effective scale $Q_{*}$ up to LL, NLL, N$^{2}$LL, and N$^{3}$LL accuracies, respectively. They are determined numerically from Eq.\eqref{per.series of Qs} and are presented as the intersections.}
\label{Exact solution of Qs}
\end{figure}

Following the standard procedures of PMC$_{\infty}$-s, the above total decay width can be rewritten as
\begin{align}\label{PMCsseries}
	\Gamma_{\rm H}|_{\text{PMC}_{\infty}\text{-s}}=\frac{M_{H}^{3}G_{F}}{36\sqrt{2}\pi}\sum_{i=1}^{5} C_{i,\rm IC}a_{s}^{i+1}(Q_{*}),
\end{align}
where $C_{i\in[1,5],\rm IC}$ are iCF coefficients which can be calculated by using Eqs.\eqref{L1}-\eqref{L5} listed in the Appendix, and the effective scale $Q_{*}$ can be determined up to N$^3$LL accuracy using the known N$^4$LO pQCD prediction,
\begin{eqnarray}	
\ln\frac{Q_{*}^{2}}{M_{H}^{2}} &=& S_{0}+S_{1}a_{s}(Q_{*}) +S_{2}a_{s}^{2}(Q_{*})+S_{3}a_{s}^{3}(Q_{*}), \label{PMCscaleExp} \\
	&=& -1.833-6.780 a_{s}(Q_{*})-906.753 a_{s}^{2}(Q_{*}) \nonumber \\
	& &-23279.302 a_{s}^{3}(Q_{*}), \label{per.series of Qs}
\end{eqnarray}
where all input parameters have been set to be their central values. As shown by Fig.\ref{Exact solution of Qs}, $Q_{*}$ can be fixed up to LL, NLL, N$^{2}$LL and N$^{3}$LL accuracies when $\Gamma_{\rm H}$ has been known up to NLO, N$^{2}$LO, N$^{3}$LO and N$^{4}$LO levels, respectively. Numerically, we have
\begin{displaymath}
	Q_{*}^{\rm LL, NLL, N^2LL, N^3LL}=\{50.081,48.164,45.266,44.407\}~{\rm GeV}.
\end{displaymath}

\begin{figure*}[htbp]
\centering
\includegraphics[width=0.45\textwidth]{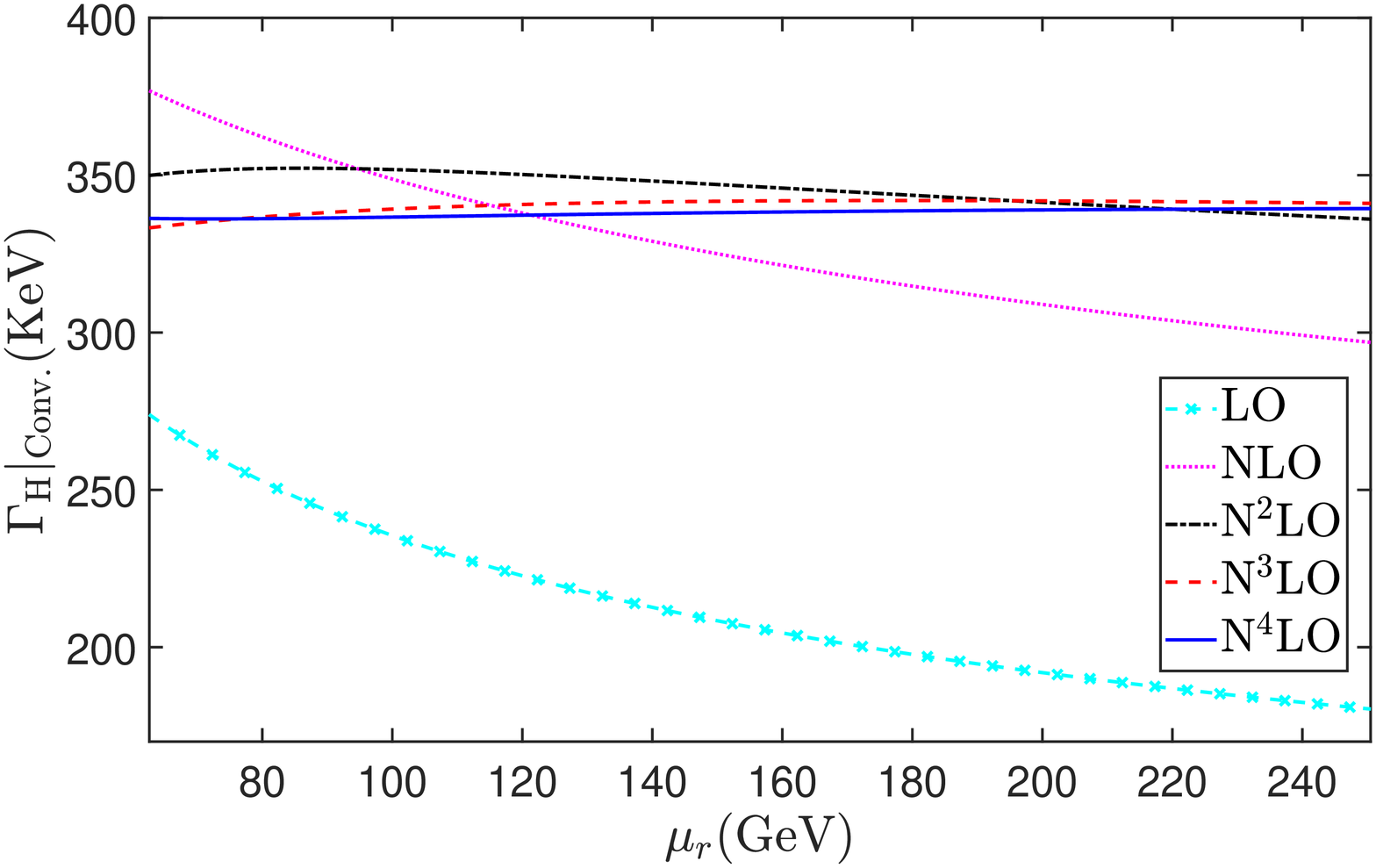}
\includegraphics[width=0.45\textwidth]{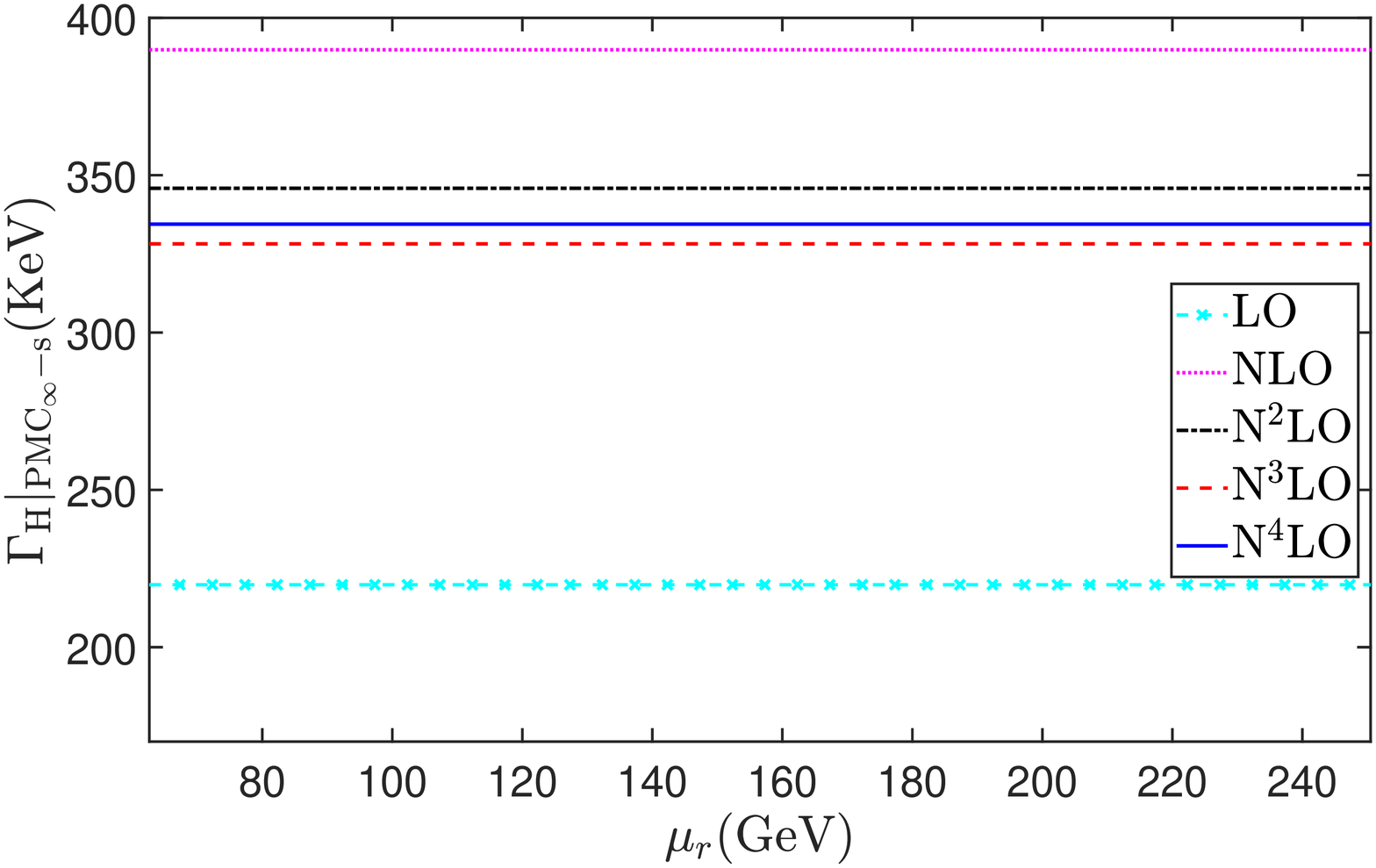}
\caption{Total decay width $\Gamma_{\rm H}$ with different QCD corrections under conventional (Upper diagram) and PMC$_{\infty}$-s (Lower diagram) scale-setting approaches, respectively. The dashed line with cross symbols, the dotted line, the dash-dot line, the dashed line, and the solid line represent the LO, NLO, N$^2$LO, N$^3$LO and N$^4$LO total decay widths, respectively.}
\label{Plot1}
\end{figure*}

\begin{table*}[htbp]
\centering
\begin{tabular}{lcccccc}
\hline
			& ~~$i=1$,~LO~~ & ~~$i=2$,~NLO~~ & ~~$i=3$,~N$^{2}$LO~~ & ~~$i=4$,~N$^{3}$LO~~ & ~~$i=5$,~N$^{4}$LO~~ & ~~Total~~ \\
\hline
			~$\Gamma_{i}|_{\rm Conv.}$~ & $218.66^{-40.95}_{+43.86}$ & $116.62^{+2.08}_{-10.67}$ & $14.44^{+25.23}_{-31.36}$ & $-8.70^{+13.84}_{-7.54}$ & $-3.57^{+1.74}_{+4.37}$ & $337.44^{+1.94}_{-1.33}$\\
\hline
			~$\Gamma_{i}|_{\rm PMC_{\infty}-s}$~ & $313.53$ & $97.16$ & $-54.18$ & $-26.42$ & $4.36$ & $334.45$\\
\hline
\end{tabular}
\caption{Total and individual decay widths (in unit: KeV) of the decay $H\to gg$ under conventional and PMC$_{\infty}$-s scale-setting approaches. As for conventional predictions, their central values are for $\mu_{r}=M_{H}$ and the errors are for $\mu_{r}\in[M_{H}/2, 2M_{H}]$. } \label{Htogg}
\end{table*}

We present the total decay width versus the renormalization scale $\mu_{r}$ before and after applying the PMC$_{\infty}$-s in Fig.~\ref{Plot1}, and the corresponding values for the total and individual decay widths of the decay $H\to gg$ up to N$^4$LO QCD corrections are given in TABLE~\ref{Htogg}. Fig.~\ref{Plot1} and TABLE~\ref{Htogg} show that, as expected, the renormalization scale dependence under conventional scale-setting approach becomes smaller-and-smaller when more-and-more loop terms have been known. For example, we obtain $\Gamma_{\rm H}|_{\rm Conv.}=337.44^{+1.94}_{-1.33}~{\rm KeV}$ for the N$^4$LO-level prediction under the choice of $\mu_{r}\in[M_{H}/2, 2M_{H}]$, whose scale error is less than $1.0\%$. Such small net scale dependence for the N$^{4}$LO prediction is due to good convergence of perturbative series and the cancellation of the scale dependence among different orders. It is found that the scale dependence for each loop terms are still large, thus the convergent behavior of the scale-invariant PMCs series can be treated as the intrinsic perturbative nature of the pQCD approximant.

\subsection{An estimation of contributions from the uncalculated N$^5$LO terms}

At present, remarkable progresses have been achieved in doing higher-order calculations in perturbation theory. However due to the complexity of loop calculations, most of perturbatively calculable high-energy observables have only been calculated at lower-orders such as NLO, NNLO and etc. Thus it is important to have a way to estimate the possible contributions from the unknown higher-order terms (UHO-terms) such that to improve the predictive power of perturbative theory.

It has been conventional to take $\mu_r$ as the typical momentum flow ($Q$) of the process to obtain the central value of the pQCD series and to then vary $\mu_r$ within a certain range such as $[Q/2, 2Q]$ as a measure of a combined effect of scale uncertainties and the contributions from the UHO terms. The shortcomings of this treatment are apparent: 1) It's effectiveness heavily depends on the convergence of series which however usually will be diluted by the divergent renormalon terms; 2) Each term in the perturbative series is highly scale-dependent and the net prediction does not satisfy the requirement of RGI; 3) One only partly obtains the information of $\{\beta_i\}$-dependent UHO-terms which control the running of $\alpha_s$ and no information on the contributions from the conformal $\{\beta_i\}$-independent terms. For the more convergent and scale-invariant PMC$_\infty$-s series, it is expected that a much better prediction of UHO contributions can be achieved. For the purpose, we need to estimate the magnitude of the UHO-terms in the perturbative series of the pQCD approximant. And we also need to know the magnitude of the UHO-terms in the perturbative series of the PMC scale such that to have an estimation of the \textit{first kind of residual scale dependence}.

Then the total uncertainty of a pQCD approximant due to the UHO-terms can be treated as the squared average of the predicted conventional scale dependence (or the \textit{first residual scale dependence}) and the predicted magnitude of the UHO-terms in the perturbative series of the pQCD approximant.

In the following, we will first try two representative approaches to estimate the magnitude of the uncalculated N$^5$LO-terms for the total decay width $\Gamma_{\rm H}$ by using the known N$^4$LO-level conventional and PMC$_\infty$-s series, respectively. The first approach is to directly predict the magnitude of the N$^5$LO-order UHO coefficient by using a fractional generating function whose parameters can be fixed by matching to the known N$^4$LO-order series, which is usually called as the Pad$\acute{\rm e}$ approximation approach (PAA)~\cite{Basdevant:1972fe, Samuel:1992qg, Samuel:1995jc}. And the second approach is to quantify the UHO's contribution in terms of a probability distribution whose representative treatment is to use the Bayes' theorem, which is called as the Bayesian approach~\cite{Cacciari:2011ze, Bagnaschi:2014wea, Bonvini:2020xeo, Duhr:2021mfd}. And then, we shall provide an estimation of the total uncertainty due to the UHO-terms.

\subsubsection{Estimation of N$^5$LO contributions using the Pad$\acute{e}$ approximation approach}

The PAA provides a systematic procedure for promoting a finite Taylor series to an analytic function. For the present known pQCD series (\ref{Convseries}) or (\ref{PMCsseries}), we can rewrite it as $\rho=a_s^2 \sum\limits_{i=1}^{5} r_{i} a_{s}^{i-1}$, where $r_{i}=\frac{M_{H}^{3}G_{F}}{36\sqrt{2}\pi} C_{i}$ or $\frac{M_{H}^{3}G_{F}}{36\sqrt{2}\pi}  C_{i, {\rm IC}}$, respectively. Its $[N/M]$-type fractional generating function is constructed as~\cite{Basdevant:1972fe, Samuel:1992qg, Samuel:1995jc}
\begin{eqnarray}
\rho^{[N/M]}(Q)  &=& a_s^2 \frac{b_0+b_1 a_s + \cdots + b_N a^N_s}{1 + c_1 a_s + \cdots + c_M a^M_s}   \label{PAAseries0} \\
                               &=& \sum_{i=1}^{5} r_{i} a^{i+1}_s + r_{6} a^{7}_s+\cdots,  \label{PAAseries}
\end{eqnarray}
where $M\geq 1$ and $N+M=4$. The coefficients $b_{i\in[0,N]}$ and $c_{i\in[1,M]}$ can be expressed by using the coefficients $r_{i\in[1,5]}$. Thus the predicted N$^{5}$LO-coefficients $r_6$ under various $[N/M]$-types are
\begin{align}
r_{6}^{[3/1]}=&\,\frac{r_5^2}{r_4},\notag\\
r_{6}^{[2/2]}=&\,\frac{-r_4^3+2 r_3 r_5 r_4-r_2 r_5^2}{r_3^2-r_2 r_4},\notag\\
r_{6}^{[1/3]}=&\,\frac{1}{r_2^3-2 r_1 r_3 r_2+r_1^2 r_4}\big(r_3^4-3 r_2 r_4 r_3^2-2 r_1 r_5
r_3^2\notag \\
	&\,+2 r_1 r_4^2 r_3+2 r_2^2 r_5 r_3+r_2^2 r_4^2+r_1^2 r_5^2-2 r_1 r_2 r_4 r_5\big), \notag\\
r_{6}^{[0/4]}=&\,\frac{1}{r_1^4}\big(-r_2^5+4 r_1 r_3 r_2^3-3 r_1^2 r_4 r_2^2-3 r_1^2 r_3^2 r_2\notag\\
	&\,+2 r_1^3 r_5 r_2+2 r_1^3 r_3 r_4\big).
\end{align}
It has been observed that the diagonal [2/2]-type Pad$\acute{\rm e}$ series is preferable for estimating the unknown contributions from the conventional pQCD series~\cite{Gardi:1996iq, Cvetic:1997qm}; while the [0/4]-type one is preferable for the PMC series~~\cite{Du:2018dma}, which makes the PAA geometric series be self-consistent with the GM-L prediction~\cite{GellMann:1954fq}.

Numerically, we obtain $C^{[2/2]}_{6}=\left(4.825^{+95.950}_{-229.971}\right)\times 10^7$ and $C^{[0/4]}_{6,{\rm IC}}\equiv 7.919\times 10^8$ for $\mu_r\in[M_H/2, 2M_H]$. Then, the magnitudes of the ${\rm N^{5}LO}$-level UHO-terms for the perturbative series (\ref{Convseries}) or (\ref{PMCsseries}) are
\begin{eqnarray}
\Delta\Gamma_{\rm H}\big|^{\rm N^{5}LO}_{\rm Conv.} &=& \pm \left|\frac{M_{H}^{3}G_{F}}{36\sqrt{2}\pi} C_{6}(\mu_r)a^7_s(\mu_r)\right|_{\rm MAX} \nonumber \\
&=& \pm2.32~{\rm KeV}, \label{UHO-conv} \\
\Delta\Gamma_{\rm H}\big|^{\rm N^{5}LO}_{\text{PMC$_\infty$-s}} &=& \pm \left|\frac{M_{H}^{3}G_{F}}{36\sqrt{2}\pi} C_{6,{\rm IC}}(Q_*)a^7_s(Q_*)\right| \nonumber \\
&=& \pm3.39~{\rm KeV}, \label{UHO-pmc}
\end{eqnarray}
where the subscript ``MAX" stands for the maximum value within the chosen $\mu_r$-region. Since $\mu_r$ can be chosen arbitrarily, the scale-invariant PMC$_\infty$-s series is a more accurate basis than the conventional series for estimating the UHO-contributions.

To estimate the \textit{first kind of residual scale dependence} for the PMC$_\infty$-s prediction (\ref{PMCsseries}), we first predict the unknown N$^4$LL-terms of the perturbative series (\ref{per.series of Qs}) by using the same procedures of PAA, whose magnitude is $\pm \left|S_4^{[0/3]} a^4_s(Q^{\rm N^{3}LL}_{*})\right|$ with $S_4^{[0/3]}=-5.838\times 10^5$. It leads to a scale shift, $\Delta Q_{*} = \pm 0.253~{\rm GeV}$, then the \textit{first kind of residual scale dependence} becomes
\begin{equation}
\Delta\Gamma_{\rm H}\big|_{\rm PMC_\infty-s}^{{\rm N^5LO},\Delta Q_{*}}=\pm 0.65~{\rm KeV}, \label{firstresid}
\end{equation}
which is smaller than the above derived conventional scale dependence,
\begin{eqnarray}
&& \Delta\Gamma_{\rm H}\big|^{{\rm N^5LO},\Delta\mu_r}_{\rm Conv.} = \left(^{+1.94}_{-1.33}\right) {\rm KeV}, \mu_{r}\in[M_{H}/2,2M_{H}], \label{Conv2scale}
\end{eqnarray}

By taking the above conventional scale dependence (\ref{Conv2scale}), or the \textit{first residual scale dependence} (\ref{firstresid}), together with the predicted magnitudes of the UHO-terms (\ref{UHO-conv}, \ref{UHO-pmc}) in the perturbative series of the pQCD approximant into consideration, the total uncertainty caused by the UHO-terms for the conventional series is
\begin{eqnarray}
\Delta\Gamma_{\rm H}\big|_{\rm Conv.}^{\rm UHO} &=& \left(^{+3.02}_{-2.67}\right)~{\rm KeV}, \label{PAAconv1}
\end{eqnarray}
and for the PMC$_\infty$-s series, it becomes
\begin{equation}
\Delta\Gamma_{\rm H}\big|_{\text{PMC$_\infty$-s}}^{\rm UHO}=\pm~3.45~{\rm KeV}. \label{PAApmc}
\end{equation}

\subsubsection{Estimation of N$^5$LO contributions using the Bayesian approach}

The Bayesian approach (B.A.) quantifies the contributions of the UHO-terms in terms of the probability distribution. The B.A. is a powerful method to construct probability distributions in which the Bayes' theorem is applied to iteratively update the probability as new information becomes available. A detailed introduction of the B.A. and its combination with the PMC approach have been given in Ref.\cite{Shen:2022nyr}, so we will only present the results here, and the interesting readers may turn to Ref.\cite{Shen:2022nyr} for all the B.A. formulas.
	
To apply the B.A., we first transform the perturbation series (\ref{Convseries}), (\ref{PMCsseries}) and (\ref{PMCscaleExp}) over $a_s=\alpha_s/4\pi$ back to the ones over $\alpha_{s}$:
\begin{eqnarray}
\Gamma_{\rm H}|_{\rm Conv.} &=& \sum_{i=1}^{5}r_{i}(\mu_{r}) \alpha_{s}^{i+1}(\mu_{r}), \\
\Gamma_{\rm H}|_{\text{PMC}_{\infty}\text{-s}} &=& \sum_{i=1}^{5}r_{i,\rm IC}\alpha_{s}^{i+1}(Q_{*}),  \\
\ln \frac{Q_{*}^{2}}{Q^{2}} &=& \sum_{i=0}^{3}\hat{S}_{i}\alpha_{s}^{i}(Q_{*}),
\end{eqnarray}
where
\begin{align}
r_{i}(\mu_{r})&=\frac{1}{(4\pi)^{i+1}} \frac{M_{H}^{3}G_{F}}{36\sqrt{2}\pi}C_{i}(\mu_{r}),\\
r_{i,\rm IC}&=\frac{1}{(4\pi)^{i+1}} \frac{M_{H}^{3}G_{F}}{36\sqrt{2}\pi}C_{i,\rm IC},\\
\hat{S}_{i}&=\frac{1}{(4\pi)^{i}}S_{i}.
\end{align}

Because the known coefficients of the conventional pQCD series are scale-dependent at every orders, the B.A. can only be applied after one specifies the choices for the renormalization scale, thus introducing extra uncertainties for the B.A. On the other hand, the PMC$_\infty$-s conformal series is scale-independent, which then provides a more reliable basis for obtaining constraints on the predictions for the UHO contributions.

Following the standard B.A. procedures, we obtain the smallest $95.5\%$ credible intervals (CIs) for the N$^5$LO coefficients $r_{6}(\mu_{r}\equiv M_H)$ and $r_{6,\rm IC}$, which are $r_{6}(\mu_{r}=M_H)\in[-1.3569,1.3569]$ and $r_{6,\rm IC}\in [-0.5624,0.5624]$, respectively. Then, the error of $\Gamma_{\rm H}$ caused by the UHO-terms for the conventional series under the B.A. is
\begin{align}
	\Delta \Gamma_{\rm H}\big|_{\rm Conv.}^{\rm N^{5}LO}=\pm 1.44~{\rm KeV}.\label{BACONVUHO}
\end{align}
If taking $\mu_{r}\in[M_H/2, 2M_H]$, we have
\begin{align}
	\Delta \Gamma_{\rm H}\big|_{\rm Conv.}^{\rm N^{5}LO}=\left(^{+2.53}_{-2.63}\right)~{\rm KeV}. \label{PMCBA2scale}
\end{align}
And for the PMC$_{\infty}$-s series, the error of $\Gamma_{\rm H}$ caused by the UHO-terms is scale-invariant and smaller,
\begin{equation}
\Delta\Gamma_{\rm H}\big|_{\text{PMC}_{\infty}\text{-s}}^{\rm N^{5}LO} = \pm 1.30~{\rm KeV}. \label{BAPMCUHO}
\end{equation}

Secondly, we observe that the predicted smallest $95.5\%$ CIs for the N$^4$LL-level coefficient $\hat{S}_{4}$ of $\ln Q_{*}^{2}/Q^{2}$ is $\hat{S}_{4}\in [-17.0331,17.0331]$, which leads to a scale shift $\Delta Q_{*} = \left(^{+0.247}_{-0.248}\right)~{\rm GeV}$ to the N$^3$LL-level scale $Q_*^{\rm N^{3}LL}=44.407$ GeV. Then the \textit{first kind of residual scale dependence} of the PMC$_\infty$-s series under the B.A. becomes
\begin{eqnarray}
\Delta \Gamma_{\rm H}\big|_{\rm PMC_{\infty}-s}^{\Delta Q_{*}} &=& \left(_{-0.63}^{+0.64}\right)~{\rm KeV}. \label{PMCBAscale}
\end{eqnarray}

\begin{figure}[htb]
\centering
\includegraphics[width=0.48\textwidth]{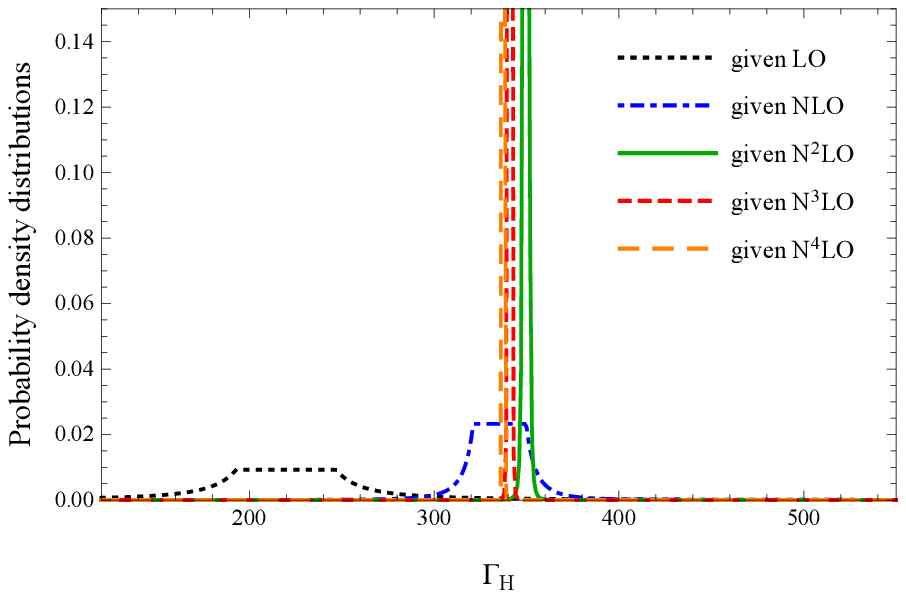}
\includegraphics[width=0.48\textwidth]{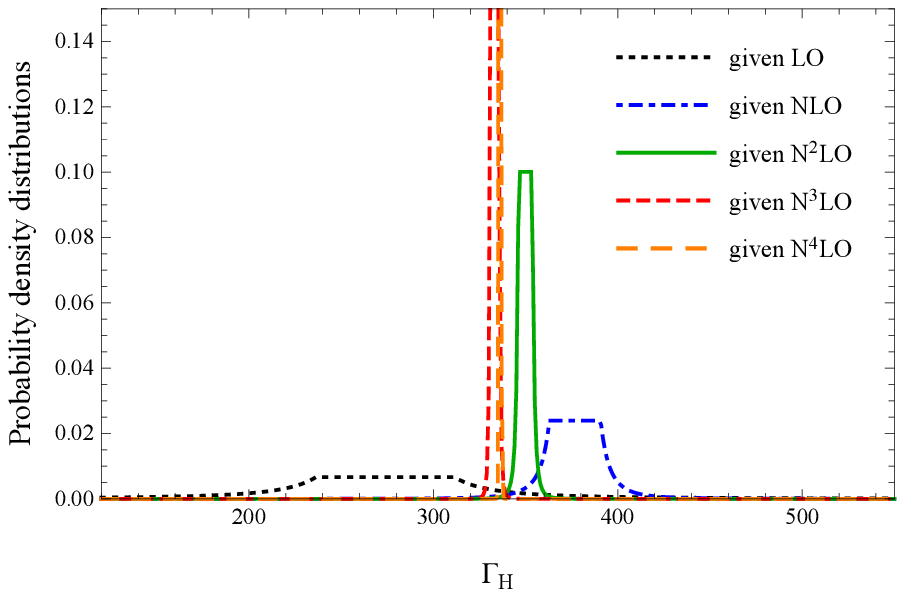}
\includegraphics[width=0.48\textwidth]{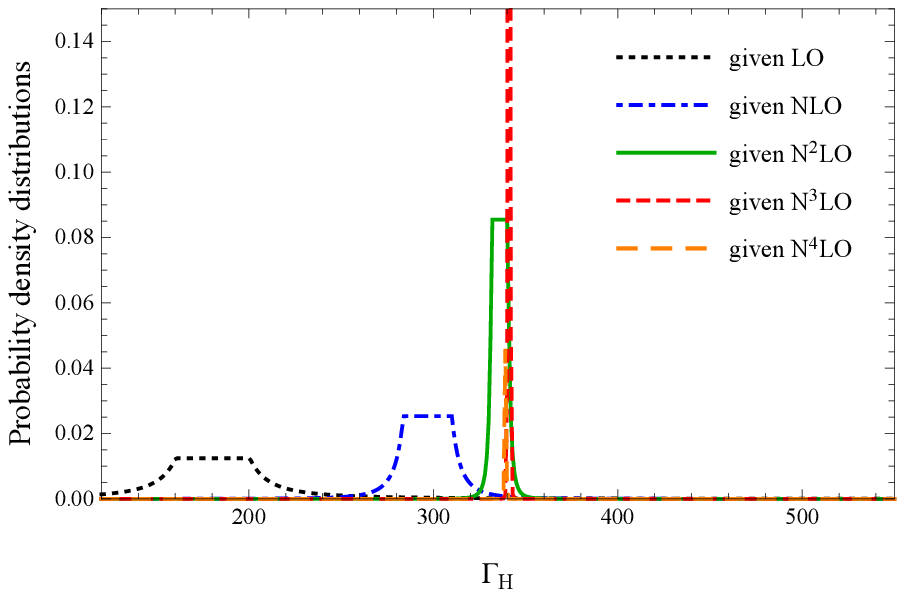}
\caption{(Color online) Probability density distributions of the decay width $\Gamma_{\rm H}$ (in unit: KeV) with different states of knowledge predicted by the conventional series with the Bayesian approach and $\mu_r=M_{H},M_{H}/2,2M_{H}$, respectively. The black dotted, the blue dash-dotted, the green solid, the red short-dashed and the orange long-dashed lines are results for the given LO, NLO, N$^2$LO, N$^3$LO and N$^4$LO series, respectively.}
\label{frhoPlot1}
\end{figure}

\begin{figure}[htb]
\centering
\includegraphics[width=0.48\textwidth]{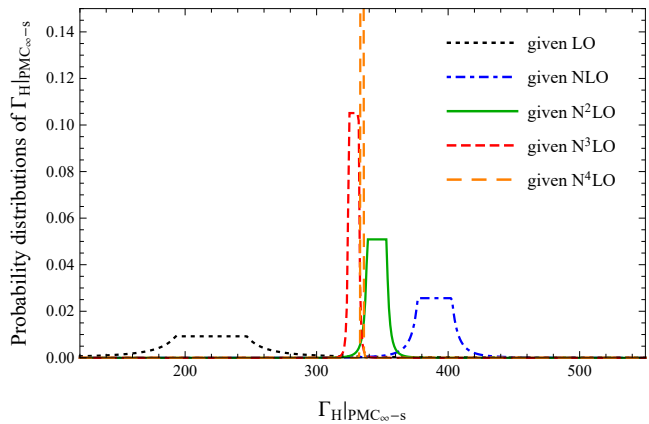}
\caption{(Color online) Probability density distributions of the decay width $\Gamma_{\rm H}$ (in unit: KeV) with different states of knowledge predicted by the PMC$_{\infty}$-s series and the Bayesian approach, respectively. The black dotted, the blue dash-dotted, the green solid, the red short-dashed and the orange long-dashed lines are results for the given LO, NLO, N$^2$LO, N$^3$LO and N$^4$LO series, respectively.}
\label{frhoPlot2}
\end{figure}

Finally, by taking the above conventional scale dependence (\ref{Conv2scale}), or the \textit{first residual scale dependence} (\ref{PMCBAscale}), and the predicted magnitudes of the UHO-terms ( \ref{PMCBA2scale}, \ref{BAPMCUHO} ) in the perturbative series of the pQCD approximant into consideration, the total uncertainty caused by the UHO-terms of the conventional series using B.A. is
\begin{eqnarray}
\Delta\Gamma_{\rm H}\big|_{\rm Conv.}^{\rm UHO} = \left(^{+3.19}_{-2.94}\right)~{\rm KeV}, \label{BAconv1}
\end{eqnarray}
and for the PMC$_\infty$-s series, it becomes
\begin{equation}
\Delta\Gamma_{\rm H}\big|_{\text{PMC$_\infty$-s}}^{\rm UHO}=\left(^{+1.45}_{-1.44}\right)~{\rm KeV}.
\end{equation}
One may observe that the error ranges estimated by using the PAA and B.A. approaches for conventional series are close to each other. While the error range of the B.A. approach is smaller than the case of the PAA approach for the PMC$_{\infty}$-s series. To explain such difference, we present Figs. \ref{frhoPlot1} and \ref{frhoPlot2} to show the probability density distributions of the conventional and the PMC$_\infty$-s decay widths $\Gamma_{\rm H}$ with different states of knowledge predicted by the Bayesian approach, respectively. Fig.~\ref{frhoPlot1} shows the results of conventional scale-setting approach under the choice of  $\mu_r=M_H/2$, $M_H$ and $2M_H$, respectively. The conventional probability density distributions are scale dependent, which lead to a broader error range as shown by Eq.(\ref{PMCBA2scale}). As shown by Fig.~\ref{frhoPlot2}, due to a sharp probability density distribution for the present known N$^4$LO-level PMC$_\infty$-s series, the B.A. error range is less than half of the PAA one. Figs. \ref{frhoPlot1} and \ref{frhoPlot2} illustrate the characteristics of the posterior distribution: a symmetric plateau with two suppressed tails. The posterior distribution given by the Bayesian approach depends on the prior distribution, and as more and more loop terms become known, the probability shall be updated with less and less dependence on the prior; i.e., the probability density becomes increasingly concentrated (the plateau becomes narrower and narrower and the tail becomes shorter and shorter) as more and more loop terms for the distribution are determined. Eqs.(\ref{BACONVUHO}, \ref{BAPMCUHO}) show that the estimated B.A. N$^5$LO contributions from the given N$^4$LO PMC$_{\infty}$-s scale-invariant series and the conventional series at the scale $\mu_r=M_H$ are very close to each other.

\begin{figure}[htb]
\centering
\includegraphics[width=0.48\textwidth]{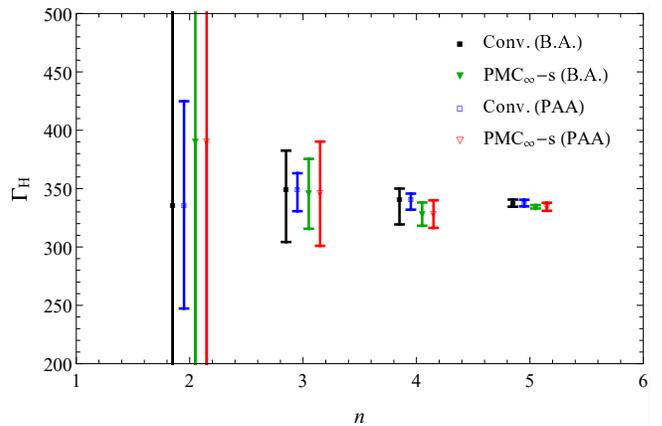}
\caption{(Color online) Comparison of conventional (Conv.) results with PMC$_\infty$-s results under B.A. and PAA, respectively. The blue hollow quadrates and the red hollow triangles represent the conventional (Conv.) and PMC$_\infty$-s predictions using PAA, respectively. And the black quadrates solid and green solid triangles represent the conventional (Conv.) and PMC$_\infty$-s predictions using B.A., respectively.}
	\label{Comparison}
\end{figure}

As a conclusion of this subsection, we present the conventional (Conv.) and PMC$_\infty$-s results up to the predicted N$^5$LO QCD corrections under the PAA and the B.A. approaches in Fig.\ref{Comparison}. Due to good convergent behavior of both series, the fixed-order pQCD predictions under the conventional and PMC$_\infty$-s scale-setting approaches are consistent with each other. For the PAA approach, because of large cancellation among the scale dependence of different orders as shown by Table~\ref{Htogg}, the conventional error bars are always smaller than the PMC$_\infty$-s ones~\footnote{Here the smaller conventional error bars are achieved for $\mu_{r}\in[M_{H}/2, 2M_{H}]$. Since $\mu_r$ is arbitrary, if taking a larger scale region, such as $\mu_{r}\in[M_{H}/3, 3M_{H}]$ or $[M_{H}/4, 4M_{H}]$, much larger error bars for conventional series shall be achieved. }. For the B.A. approach, because of sharper probability density distributions at higher-orders, the PMC$_\infty$-s error bars are always smaller than the conventional ones. The differences for both cases become smaller when more loop terms have been known.

\subsection{Uncertainties due to $\Delta M_H$, $\Delta m_t$ and $\Delta \alpha_s(M_Z)$}

In addition to the errors caused by the UHO-terms, there are also errors caused by the input parameters $M_{H}$, $m_t$ and $\alpha_s(M_Z)$, e.g. $\Delta M_{H}=\pm0.17~{\rm GeV}$, $\Delta m_{t}=\pm0.30~{\rm GeV}$, and $\Delta \alpha_{s}(M_Z)=\pm0.0009$~\cite{ParticleDataGroup:2020ssz}. When discussing the errors from one parameter, the other parameters are set to be their central value. We obtain
\begin{align}
	&\Delta\Gamma_{\rm H}\big|_{\rm Conv.}^{\Delta M_{H}}=\left(^{+1.22}_{-1.21}\right)~{\rm KeV},\\
	&\Delta\Gamma_{\rm H}\big|_{\rm Conv.}^{\Delta m_{t}}=\pm 0.01~{\rm KeV},\\
	&\Delta\Gamma_{\rm H}\big|_{\rm Conv.}^{\Delta\alpha_{s}(M_Z)}=\left(^{+6.27}_{-6.20}\right)~{\rm KeV},
\end{align}
and
\begin{align}
	&\Delta\Gamma_{\rm H}\big|_{\text{PMC}_{\infty}\text{-s}}^{\Delta M_{H}}=\pm 1.21~{\rm KeV},\\
	&\Delta\Gamma_{\rm H}\big|_{\text{PMC}_{\infty}\text{-s}}^{\Delta m_{t}}=\pm 0.02~{\rm KeV},\\
	&\Delta\Gamma_{\rm H}\big|_{\text{PMC}_{\infty}\text{-s}}^{\Delta\alpha_{s}(M_Z)}=\left(^{+6.05}_{-6.00}\right)~{\rm KeV}.
\end{align}
It shows that the magnitude of $\Delta\alpha_{s}(M_Z)$ dominates the error sources. Thus a more precise measurements on the reference point $\alpha_s(M_Z)$ is important for a more precise pQCD prediction.

By taking the error ranges caused by the UHO-terms that have been predicted via the PAA and B.A. into consideration and by adding all the mentioned errors in quadrature, our final results for the $H\to gg$ total decay width $\Gamma_{\rm H}$ using the PAA predictions are
\begin{eqnarray}
&& \Gamma_{\rm H}\big|_{\rm Conv.}^{\rm PAA} = 337.44^{+7.07}_{-6.86}~{\rm KeV}, \\
&& \Gamma_{\rm H}\big|_{\text{PMC}_{\infty}\text{-s}}^{\rm PAA} = 334.45^{+7.07}_{-7.03}~{\rm KeV},
\end{eqnarray}
whose net errors are $4.1\%$ and $4.2\%$, respectively. And the final results using the B.A. predictions are
\begin{eqnarray}
&& \Gamma_{\rm H}\big|_{\rm Conv.}^{\rm B.A.} = 337.44^{+7.14}_{-6.96}~{\rm KeV}, \\
&& \Gamma_{\rm H}\big|_{\text{PMC}_{\infty}\text{-s}}^{\rm B.A.} = 334.45^{+6.34}_{-6.29}~{\rm KeV},
\end{eqnarray}
whose net errors are $4.2\%$ and $3.8\%$, respectively.  \\

\section{Summary} \label{Summary}

In the paper, we have started from the PMC and then proposed a novel single-setting procedure by using the iCF property of the renormalizable gauge theories, e.g. the PMC$_{\infty}$-s approach, to eliminate the conventional renormalization scale ambiguities. On the one hand, when enough UHO-terms have been known, the conventional series can achieve small scale-dependent prediction due to the cancellation of scale-dependence among different orders. On the other hand, the PMC$_{\infty}$-s approach removes the scale-dependent terms by using the RGE recursively, and then achieves a fixed-order prediction free of conventional renormalization scale ambiguity. Using the Higgs decays into two gluons as an explicit example, we have shown that the PMC$_{\infty}$-s approach greatly suppresses the residual scale dependence caused by the UHO-terms, especially when using the B.A. approach to estimate the UHO contributions. To compare with the scale-dependent conventional series, the scale-invariant PMC$_{\infty}$-s series provides a more accurate and reliable platform for estimating the UHO-contributions. After applying the PMC$_{\infty}$-s procedures, the resultant conformal series also makes its prediction be scheme independent, which satisfies the requirement of the standard renormalization group invariance and can be ensured by the commensurate scale relations among different orders~\cite{Brodsky:1994eh, Huang:2020gic}.

The PMCs approach adopts all the RG-involved non-conformal $\{\beta_i\}$-terms to achieve an overall effective coupling of the process, whose argument (the PMC scale) corresponds to the correct momentum flow of the process. While the PMC$_\infty$-s approach adopts the property of iCF to fix the overall effective scale, which ensures the scale invariance of the pQCD series at each order by only identifying the $\{\beta_0\}$-terms at each order. The scale-setting procedures of the PMC$_\infty$-s approach is simpler than the PMCs, since it does not need to apply the degeneracy relations. The equivalence of those two single-scale setting approaches indicates that by using the RGE to fix the value of effective coupling is equivalent to require each loop terms satisfy the scale invariance simultaneously, and vice versa. The scale-invariant perturbative series shows the intrinsic perturbative nature of a pQCD observable. Thus the way of using RGE provides a solid way to solve the conventional scale-setting ambiguity. The PMC single-scale setting approach can be applied to any perturbative series in case that we have known the corresponding RGE and correctly applied it to fix the magnitude of the expansion parameter. \\

\noindent{\bf Acknowledgements:} This work was supported by the graduate research and innovation foundation of Chongqing, China under Grant No.ydstd1912, and by the Natural Science Foundation of China under Grant No.11905056, No.12175025 and No.12147102.

\section*{Appendix: The intrinsic conformal coefficients up to N$^{5}$LO-level and coefficients of PMC$_{\infty}$ scale up to N$^{4}$LL-level}
\label{A}

From Eq.~\eqref{relation}, the intrinsic conformal coefficients $\mathcal{L}_{i,{\rm IC}}$ and $S_{i}$ can be obtained order-by-order by fixing the the number of flavor $n_{f}=33/2$:
\begin{widetext}
\begin{align}
\mathcal{L}_{1,\rm IC}=&\,\mathcal{L}_{1}(Q),\label{L1}\\
\mathcal{L}_{2,\rm IC}=&\,\mathcal{L}_{2}(Q)\big|_{n_{f}=\frac{33}{2}},\\
\mathcal{L}_{3,\rm IC}=&\,\mathcal{L}_{3}(Q)\big|_{n_{f}=\frac{33}{2}}+n\bar{\beta}_{1}\bar{S}_{0}\mathcal{L}_{1}(Q),\\
\mathcal{L}_{4,\rm IC}=&\,\mathcal{L}_{4}(Q)\big|_{n_{f}=\frac{33}{2}} +(n+1)\bar{\beta}_{1}\bar{S}_{0}\mathcal{L}_{2}(Q)\big|_{n_{f}=\frac{33}{2}}+n\left(\bar{\beta}_{1}\bar{S}_{1}+\bar{\beta}_{2}\bar{S}_{0}\right)\mathcal{L}_{1}(Q),\\
\mathcal{L}_{5,\rm IC}=&\,\mathcal{L}_{5}(Q)|_{n_{f}=\frac{33}{2}}+(n+2)\bar{\beta}_{1}\bar{S}_{0} \mathcal{L}_{3}(Q)\big|_{n_{f}=\frac{33}{2}} +(n+1)\Big(\bar{\beta}_{1}\bar{S}_{1}+\bar{\beta}_{2}\bar{S}_{0}\Big)\mathcal{L}_{2}(Q)\big|_{n_{f}=\frac{33}{2}}+n\Big(\frac{n+2}{2}\bar{\beta}_{1}^{2}\bar{S}_{0}^{2}+\bar{\beta}_{1}\bar{S}_{2}\notag\\
		&+\bar{\beta}_{2}\bar{S}_{1}+\bar{\beta}_{3}\bar{S}_{0}\Big)\mathcal{L}_{1}(Q),\label{L5}\\
\mathcal{L}_{6,\rm IC}=&\,\mathcal{L}_{6}(Q)|_{n_{f}=\frac{33}{2}}+ (n+3)\bar{\beta}_{1}\bar{S}_{0}\mathcal{L}_{4}(Q)\big|_{n_{f}=\frac{33}{2}}+(n+2)\Big(\bar{\beta}_{1}\bar{S}_{1}+\bar{\beta}_{2}S_{0}\Big)\mathcal{L}_{3}(Q)\big|_{n_{f}=\frac{33}{2}}+(n+1)\Big(\frac{n+3}{2}\bar{\beta}_{1}^{2}S_{0}^{2}+\bar{\beta}_{1}\bar{S}_{2}\notag\\
		&+\bar{\beta}_{2}\bar{S}_{1}+\bar{\beta}_{3}\bar{S}_{0}\Big) \mathcal{L}_{2}(Q)|_{n_{f}=\frac{33}{2}}+n\Big(\frac{2n+5}{2}\bar{\beta}_{1}\bar{\beta}_{2}\bar{S}_{0}^{2} +(n+2)\bar{\beta}_{1}^{2}\bar{S}_{0}\bar{S}_{1}+\bar{\beta}_{1}\bar{S}_{3} +\bar{\beta}_{2}\bar{S}_{2}+\bar{\beta}_{3}\bar{S}_{1}+\bar{\beta}_{4}\bar{S}_{0}\Big)\mathcal{L}_{1}(Q),\\
S_{0}=&\,\frac{1}{n\beta_{0}\mathcal{L}_{1}(Q)}\left[\mathcal{L}_{2,\rm IC}-\mathcal{L}_{2}(Q)\right],\label{S0}\\
S_{1}=&\,\frac{1}{n\beta_{0}\mathcal{L}_{1}(Q)}\bigg[\mathcal{L}_{3,\rm IC}-\mathcal{L}_{3}(Q)-(n+1)\beta_{0}S_{0}\mathcal{L}_{2}(Q)-n\left(\frac{n+1}{2}\beta_{0}^{2}S_{0}^{2}+\beta_{1}S_{0}\right)\mathcal{L}_{1}(Q)\bigg],\\
S_{2}=&\,\frac{1}{n\beta_{0}\mathcal{L}_{1}(Q)}\bigg[\mathcal{L}_{4,\rm IC}-\mathcal{L}_{4}(Q) -(n+2)\beta_{0}S_{0}\mathcal{L}_{3}(Q)-(n+1)\bigg(\frac{n+2}{2}\beta_{0}^{2}S_{0}^{2} +\beta_{0}S_{1}+\beta_{1}S_{0}\bigg)\mathcal{L}_{2}(Q)\notag\\
		&-n\bigg(\frac{(n+1)(n+2)}{3!}\beta_{0}^{3}S_{0}^{3} +(n+1)\beta_{0}^{2}S_{0}S_{1} +\frac{2n+3}{2}\beta_{0}\beta_{1}S_{0}^{2} +\beta_{1}S_{1}+\beta_{2}S_{0}\bigg)\mathcal{L}_{1}(Q)\bigg]\\
S_{3}=&\,\frac{1}{n\beta_{0}\mathcal{L}_{1}(Q)}\Big[\mathcal{L}_{5,\rm IC}-\mathcal{L}_{5}(Q)-(n+3)\beta_{0}S_{0}\mathcal{L}_{4}(Q) -(n+2)\left(\frac{n+3}{2}\beta_{0}^{2}S_{0}^{2}+\beta_{0}S_{1}+\beta_{1}S_{0}\right)\mathcal{L}_{3}(Q)\notag\\
		&-(n+1)\Big(\frac{(n+2)(n+3)}{6}\beta_{0}^{3}S_{0}^{3} +(n+2)\beta_{0}^{2}S_{0}S_{1}+\frac{2n+5}{2}\beta_{0}\beta_{1}S_{0}^{2}+\beta_{0}S_{2}+\beta_{1}S_{1}+\beta_{2}S_{0}\Big)\mathcal{L}_{2}(Q)\notag\\
		&-n\Big(\frac{(n+1)(n+2)(n+3)}{4!}\beta_{0}^{4}S_{0}^{4} +\frac{(n+1)(n+2)}{2}\beta_{0}^{3}S_{0}^{2}S_{1}+\frac{n+1}{2}\beta_{0}^{2}S_{1}^{2}+(n+1)\beta_{0}^{2}S_{0}S_{2}\notag\\
		&\,+\frac{3n^{2}+12n+11}{6}\beta_{0}^{2}\beta_{1}S_{0}^{3} +(2n+3)\beta_{0}\beta_{1}S_{0}S_{1}+(n+2)\beta_{0}\beta_{2}S_{0}^{2} +\frac{n+2}{2}\beta_{1}^{2}S_{0}^{2}+\beta_{1}S_{2}+\beta_{2}S_{1}+\beta_{3}S_{0}\Big)\mathcal{L}_{1}(Q)\Big].\label{S3}\\
S_{4}=&\,\frac{1}{n\beta_{0}\mathcal{L}_{1}(Q)}\Bigg[\mathcal{L}_{6,{\rm IC}}-\mathcal{L}_{6}(Q)-(n+4)\beta_0S_0\mathcal{L}_{5}(Q) -(n+3) \left(\frac{n+4}{2}\beta_0^2S_0^2+\beta_0S_1+\beta_1S_0\right)\mathcal{L}_{4}(Q)\notag\\
		&-(n+2) \bigg(\frac{(n+3)(n+4)}{3!}\beta_0^3S_0^3+(n+3)\beta_0^2S_0S_1 +\frac{2n+7}{2}\beta_0\beta_1S_0^2+\beta_0S_2+\beta_1S_1+\beta_2S_0\bigg)\mathcal{L}_{3}(Q)\notag\\
		&-(n+1)\bigg(\frac{(n+2)(n+3)(n+4)}{4!} \beta_0^4S_0^4+\frac{(n+2)(n+3)}{2}\beta_0^3S_0^2S_1 +(n+2)\beta_0^2\left(\frac{1}{2}S_1^2+S_0S_2\right)+(2n+5)S_0S_1\beta_0\beta_1\notag\\
		&+(n+3)\left(\frac{1}{2}\beta_1^2+\beta_0\beta_2\right)S_0^2 +\frac{3n^2+18n+26}{6}S_0^3\beta_0^2\beta_1+\beta_0S_3+\beta_1S_2+\beta_2S_1+\beta_3S_0\bigg)\mathcal{L}_{2}(Q)\notag\\
		&-n\bigg(\frac{(n+1)(n+2)(n+3)(n+4)}{5!}\beta_0^5S_0^5 +\frac{(n+1)(n+2)(n+3)}{6}\beta_0^4S_0^3S_1+\frac{(n+1)(n+2)}{2}\beta_0^3\left(S_0S_1^2+S_0^2S_2\right)\notag\\
		&+(n+1)\beta_0^2\left(S_1S_2+S_0S_3\right)+(n+2)\beta_1^2S_0S_1 +\frac{3n^2+15n+17}{6}\beta_0\beta_1^2S_0^3+\frac{2n+3}{2}\beta_0\beta_1S_1^2+(2n+3)\beta_0\beta_1S_0S_2\notag\\
		&+\frac{3n^2+12n+11}{2}\beta_0^2\beta_1S_0^2S_1+\frac{(2n+5)(n^2+5n+5)}{12} \beta_0^3\beta_1S_0^4+2(n+2)\beta_0\beta_2S_0S_1+\frac{(n+2)(n+3)}{2}\beta_0^2\beta_2S_0^3\notag\\
		&+\frac{2n+5}{2}\left(\beta_0\beta_3+\beta_1\beta_2\right)S_0^2+\beta_1S_3 +\beta_2S_2+\beta_3S_1+\beta_4S_0\bigg)\mathcal{L}_{1}(Q)\Bigg]\label{S4},
	\end{align}
\end{widetext}
where $\bar{\beta}_{i}=\beta_{i}|_{n_{f}=33/2}$ and $\bar{S}_{i}=S_{i}|_{n_{f}\to 33/2}$. In the above equations ${\cal L}(Q)$ is a short notation for ${\cal L}(Q, Q)$. It is not hard to find that $\bar{S}_{i}$ are $0/0$-type limits which always converge to finite values because there are always components in the numerator that can always cancel with the $\beta_{0}$ in the denominator.

Then the relations between $\{S_{i}\}$ and $\{F_{i}\}$ can be calculated iteratively as follows:
\begin{align}
	F_{0}=&\,S_{0},\\
	F_{1}=&\,S_{1},\\
	F_{2}=&\,S_{2}-\beta_{0}\hat{S}_{0}S_{1}\label{T2},\\
	F_{3}=&\,S_{3}-2\beta_{0}\hat{S}_{0}S_{2}+\left(\beta_{0}^{2}\hat{S}_{0}^{2}-\beta_{1}\hat{S}_{0}\right)S_{1}-\beta_{0}S_{1}^{2}\label{F3},\\
	F_{4}=&\,S_{4}-2\beta_{1}\hat{S}_{0}S_{2}-\beta_{2}\hat{S}_{0}S_{1}-\beta_{0}^{3}\hat{S}_{0}^{3}S_{1}-\beta_{1}S_{1}^{2}\notag\\
	&\,+3\beta_{0}^{2}\left(\hat{S}_{0}S_{1}^{2}+\hat{S}_{0}^{2}S_{2}\right)-3\beta_{0}\left(\hat{S}_{0}S_{3}+S_{1}S_{2}\right)\notag\\
	&\,+\frac{5}{2}\beta_{0}\beta_{1}\hat{S}_{0}^{2}S_{1}\label{F4},
\end{align}
where $\hat{S}_{0}=S_{0}-\ln Q_{0}^{2}/Q^{2}$.

\end{document}